\documentclass[alpha-refsdoi]{wiley-article}
\usepackage{color}
\usepackage{ulem}
\usepackage{amsmath}
\usepackage{siunitx}
\usepackage{parskip}
\usepackage{times}
\usepackage{caption}
\usepackage{subcaption}
\usepackage{comment}
\usepackage{xurl}
\usepackage{hyperref}
\usepackage{xr}
\usepackage{cleveref}
\usepackage{lineno}
\usepackage{cancel}
\papertype{Original Article}
\paperfield{Journal Section}
\DeclareMathOperator{\sgn}{sgn}

\title{A Moisture–Vorticity Theory for the Boreal Summer Quasi-Biweekly Oscillation}

\abbrevs{QBWO, Quasi-biweekly Oscillation; OLR, Outgoing Longwave Radiation; JJAS, June-July-August-September; WTG, Weak Temperature Gradient.}

\author[1\authfn{1}\authfn{2}]{Shubhrangshu Biswas}
\author[1,2\authfn{1}\authfn{2}]{Jai Sukhatme}
\author[1,3\authfn{1}\authfn{2}]{Bishakhdatta Gayen}
\contrib[\authfn{1}]{Equally contributing authors.}

\affil[1]{Centre for Atmospheric and Oceanic Sciences, Indian Institute of Science, Bengaluru, Karnataka-560012, India}
\affil[2]{Divecha Centre for Climate Change, Indian Institute of Science, Bengaluru, Karnataka-560012, India}
\affil[3]{Mechanical Engineering Department, University of Melbourne, Melbourne, VIC-3010, Australia}

\corraddress{Shubhrangshu Biswas, Centre for Atmospheric and Oceanic Sciences, Indian Institute of Science, Bengaluru, Karnataka-560012, India}
\corremail{shubhrangshu@iisc.ac.in}

\presentadd[\authfn{2}]{Centre for Atmospheric and Oceanic Sciences, Indian Institute of Science, Bengaluru, Karnataka-560012, India}

\fundinginfo{Shubhrangshu Biswas acknowledges Prime Minister's Research Fellowship, Indian Institute of Science and Grantham Fellowship for supporting his Ph.D.}

\runningauthor{Biswas et al.}

\begin{document}
\maketitle

\begin{abstract}

We develop a moisture vorticity theory for the boreal summer quasi-biweekly oscillation (QBWO) in the tropics and analyze it with the aid of GPT-5.5. The model is formulated on a weak-temperature-gradient slow manifold, in which moisture anomalies diagnose the divergent circulation while the rotational circulation evolves through Rossby wave dynamics and vortex stretching. %Rather than treating relatively dry and very moist regions as distinct dynamical systems, we show that both arise as regional limits of a common coupled moisture–vorticity framework. 
The key coupling is controlled by eddy advection of the background moisture gradient, with the projection $(\mathbf{k}\cdot\nabla\bar q)$  modifying propagation and the cross-gradient factor  $(\mathbf{k}\times\nabla\bar q)_z$  controlling moisture extraction and growth. In the weak-coupling regime, which is relevant for the real-world situation, the QBWO is a Doppler-shifted Rossby mode whose growth or decay is determined by phase-coherent moisture extraction and subsequent vortex stretching; moisture damping weakens this feedback but does not introduce a sharp threshold in the minimal linear model. In terms of subtropical geographical locations, this renders the QBWO unstable over the Bay of Bengal with an intraseasonal growth rate and co-located vorticity and moisture anomalies, and it decays over relatively drier regions such as Central and West Africa with a quadrature relation between vorticity and moisture anomalies. The equations also admit a strong-coupling regime where the system approaches a moist-vortex limit. The theory interprets the QBWO as a Rossby-moisture–vorticity instability whose regional behavior depends on background moisture-gradient geometry, coupling strength, damping, and mean-flow advection.

%Using a reduced moisture–vorticity framework, with the aid of GPT-5.5 we examine the quasi-biweekly oscillation (QBWO) over relatively dry and very moist tropical regions. The analysis is framed on a weak-temperature-gradient approximation appropriate to slow, large-scale tropical dynamics. Specifically, moisture anomalies evolve and are used to diagnose the divergent circulation, which in turn act as a forcing for evolving the vorticity anomalies. In dry regions, the system reduces to a Doppler-shifted Rossby mode weakly coupled to moisture anomalies. The background meridional moisture gradient modifies the real frequency and introduces an imaginary component; for poleward-decreasing moisture, this favors decay and a quadrature relationship between vorticity and moisture/divergence anomalies. In very moist regions, eddy advection of the background moisture gradient generates moisture anomalies that feed back on the rotational gyre through divergence and vortex stretching. This yields an alignment between the QBWO gyre, convergence and moisture anomalies. The resulting dynamics suggest a moist Rossby–vortex instability in which QBWO growth is favored in very moist - large background moisture gradient regions.

\end{abstract}
%\linenumbers

\section{Introduction}
The Quasi-Biweekly Oscillation (QBWO) is observed throughout the tropical regions of the globe. In boreal summer, QBWO shows high variance over the South China Sea and Bay of Bengal basins, while in boreal winter, it shows high variance in the south-west tropical Pacific ocean \citep{kikuchi2009global}. Although showing lower activity, QBWO can be traced and tracked over other regions in the tropics, where it sets the large scale environment for different synoptic atmospheric phenomena. As its name suggests, it has a characteristic time period of 10–25 days, and plays an important role in bridging synoptic-scale disturbances and intraseasonal oscillations. Spatially, these are large-scale systems, typically associated with planetary wavenumbers of 10 or less. There is no strict constraint on their direction of propagation; impelled by different mechanisms, they can traverse eastward, westward \citep{kikuchi2009global}, and even poleward \citep{wang2017quasi,yangSH,sambrita}. Understanding the mechanisms governing their propagation, growth, and regional structure is therefore important for explaining tropical variability across a broad range of spatial and temporal scales.

The predominance of westward propagation has led QBWO disturbances to be commonly interpreted as manifestations of equatorial Rossby (ER) waves \citep{kiladis2009convectively,chen2010characteristics,kikuchi2009global}.
%Nevertheless, the majority of QBWO events propagate westward and are generally regarded as manifestations of Equatorial Rossby (ER) waves  modified by moisture coupling and the mean background flow \citep{biswas2026impact}.
ER waves arise naturally in the tropics due to the meridional variation of Earth's vertical component of rotation, commonly referred to as the $\beta$-effect. These waves emerge as one of the analytical solutions of the shallow-water equations formulated on an equatorial $\beta$-plane under the assumption of plane-wave oscillations \citep{matsuno1966quasi}. In the absence of moisture coupling and with a resting basic state, ER waves propagate westward and exhibit a symmetric structure consisting of alternating cyclonic and anticyclonic circulations about the equator for the first meridional mode ($n=1$). In this framework, horizontal convergence (divergence) is located on the eastern (western) flank of the cyclonic circulation, implying that divergence leads vorticity by a phase difference of $\pi/2$, or equivalently, is in quadrature with it.
In tropical studies, ER waves are commonly identified by filtering data within the frequency–wavenumber domain corresponding to the theoretical ER-wave dispersion relation \citep{wk1999}.
%However, caution is required, as such filtering does not always guarantee the isolation of pure ER-wave signals \citep{Kreview}. The filtered ER waves generally exhibit substantial agreement with the dry theoretical framework. Nevertheless, moisture coupling introduces important modifications to their structure and propagation characteristics. 
%Several observed features cannot be explained by dry dynamics alone \citep{kw1995,nakamura2022convective,matthews2025vorticity}, highlighting the indispensable role of convection in convectively coupled equatorial Rossby (CCER) waves.
However, such filtering does not necessarily isolate a dynamically pure ER mode \citep{Kreview}, and several observed QBWO characteristics cannot be explained by dry ER-wave dynamics alone \citep{kw1995,nakamura2022convective,matthews2025vorticity}. In particular, the relationship between moisture and rotational vorticity varies markedly among tropical regions, highlighting the indispensable role of convection in convectively coupled equatorial Rossby (CCER) waves. Over the South Asian monsoon region, moisture, convection, and cyclonic vorticity anomalies are approximately collocated, whereas over many other tropical regions, moisture and vorticity are closer to quadrature \citep{biswas2026impact}. This contrasting vortex–moisture morphology requires an explanation beyond the conventional dry shallow-water framework.

A range of moist dynamical theories has been proposed to explain the structure and propagation of QBWO and convectively coupled ER disturbances. These include evaporation–wind feedback \citep{goswami1992mechanisms}, wind-induced surface heat exchange, and cloud–radiative feedbacks \citep{fuchs2019simple,chen2022model}. Such processes can supply moisture, modify diabatic heating, alter the pressure field, and reduce the phase speed of convectively coupled ER waves. More broadly, moisture-mode theories treat moisture as a prognostic variable that actively controls convection and circulation rather than representing convection only through a reduced equivalent depth. This framework has been applied to the Madden–Julian Oscillation \citep{raymond2009moisture,adames2016mjo,adames2021moisture,kang2021role}, the Boreal Summer Intraseasonal Oscillation \citep{ajayamohan2011poleward,wang2020diagnosing,ghatak2025northward}, the QBWO \citep{gonzalez2019distinct,mayta2022westward,dong2024propagation,yangSH,sambrita}, and synoptic-scale tropical disturbances \citep{suhasboos,lahaye2016,adames2018,diaz2019,chaud}. Recent work further suggests that convectively coupled ER-like disturbances over tropical oceans can exhibit moisture-mode behaviour \citep{mayta2024stirring}. These studies motivate viewing the QBWO as an ER-like rotational disturbance whose thermodynamic structure, growth, and propagation are modified by explicit moisture dynamics. However, most existing theories emphasize particular mechanisms or geographical regions. A unified theoretical explanation for why westward-propagating QBWO disturbances exhibit nearly collocated moisture and vorticity over South Asia but an approximately quadrature relationship elsewhere remains lacking.

The observational analysis of \citep{biswas2026impact} provides the motivation for constructing such a framework. Despite strong regional differences in the mean wind, background vorticity, and moisture distribution, a common set of leading-order balances can be identified in the QBWO vorticity and moisture budgets. Across much of the tropics, westward propagation is strongly influenced by advection by the mean easterly flow, whereas over the South Asian monsoon region, eddy advection of the background vorticity and moisture gradients becomes particularly important. These regional differences are accompanied by pronounced variations in the moisture–vorticity phase relationship. Along with the equations of vorticity and moisture in the regions of study \citep{biswas2026impact}, a third relation connecting moisture and wind field is required to solve the system completely. Processes connecting thermodynamics and dynamics are knotty in general, but a simple closure relation, preferably inspired by observation, can make theoretically trackable analysis of the system.
%At the same time, column-integrated water-vapour anomalies and mid- to lower-tropospheric horizontal divergence exhibit an approximately linear spatial association across most of the tropical regions examined, with only moderate regional variations in the regression slope. We use this empirical relationship to close the otherwise incomplete vorticity–moisture system by relating anomalous moisture linearly to the divergent circulation.
This closure equation is motivated by the weak temperature gradient (WTG) approximation \citep{sobel2001weak,sobel2002water}, where the divergence scales linearly with diabatic heating. At the large spatial scales and relatively slow timescales of QBWO disturbances, rapid gravity-wave adjustment suppresses free-tropospheric temperature anomalies, leaving a leading-order thermodynamic balance between vertical advection of the background dry static energy and diabatic heating \citep{adames2019scale,adames2021moisture,snide2022role}. If diabatic heating is linearized as a function of column-moisture anomalies \citep{mayta2022westward,mayta2024stirring}; vertical motion—and, through continuity, the associated horizontal convergence—can be related approximately to moisture. The proposed closure can therefore be regarded as an observationally motivated, WTG-consistent approximation rather than an exact representation of moist convection.

In this study, we verify the applicability of a linear connection between moisture and divergence for our system. Following that, we combine the observed dominant vorticity and moisture balances with this linear closure to derive an analytically tractable moisture–vorticity model of the QBWO. We ask four principal questions: Can the reduced system reproduce westward QBWO propagation across tropical regions with substantially different background states? What controls the regional phase relationship between moisture, convergence, and vorticity? Under what orientation of the background moisture gradient relative to the disturbance shape-vector does moisture coupling produce growth or decay? How does the system change between dry Rossby waves, weak, and strong coupling regimes, defined by the relative importance of the background moisture and vorticity gradients? By deriving the corresponding dispersion relation and energy budget, we distinguish the roles of the $\beta$-effect, mean-flow advection, background-gradient advection, and moisture-induced vortex stretching. We then apply the theory to several tropical regions and compare its predictions with observations. Section 2 describes the data and analysis methods. Section 3 introduces the reduced governing equations, closure relation, and non-dimensional formulation. Section 4 presents the solutions for different coupling regimes and the energetic interpretation of wave growth and decay. Section 5 applies the theory regionally, and Section 6 summarizes the principal findings, limitations, and directions for future development. Details of a few mathematical steps related to the key results are given in the appendix in section 7.

\section{Data and Methods}

Our use of Large Language Models (LLMs; in particular we use ChatGPT 5.5) is motivated by recent work in fields of physical, computer and mathematical sciences \citep{inference1,inference2}. As noted by these authors, we found the LLM to be an extremely capable collaborator, specifically in making connections to disparate situations from literature, accelerating calculations and estimates. %We began with GPT 5.2 but this required a fair bit of supervision to avoid algebraic errors, and more importantly, making unwarranted assumptions while analyzing the equations at hand. GPT 5.5 was much improved in all these aspects. 
%%Broadly, our usage of the LLM began with providing it with the equations for vorticity and moisture --- these were formulated based on our study of QBWO budgets in reanalysis \citep{biswas2026impact}. %Remarkably, without any context the model more-or-less identified the physical relevance of the equations. 
%With clarifications such as informing the LLM that we are working on the QBWO, defining background variables, and that the set of equations govern anomalies of fields, we %were ready for their analysis. At each stage, we 
%%At each stage, we suggested a path of analysis and the LLM proceeded with calculations which we had performed manually. In cases like non-dimensionalization, appropriate scale parameters were suggested for our goals. %Finally, we also used the LLM for code generation in the initial value calculations, here its performance was largely driven by a single prompt.
%Apart from specific calculations, we also discussed the physical meaning of the results, and finally, 
The LLM also helped in improving our presentation of the results in this manuscript.

For generating the figures to compare the theory with observations, we have used the same dataset sources as mentioned in \citep{biswas2026impact}, with a few additional datasets for the energy/radiation. Twenty-one years (1990-2010) of daily mean of outgoing longwave radiation (OLR) from the National Oceanic and Atmospheric Administration (NOAA) (\url{https://www.ncei.noaa.gov/products/climate-data-records/outgoing-longwave-radiation-daily}) is used as a proxy for the convection. Wind, specific humidity, temperature, and geopotential data are obtained from ERA5 reanalysis data at different pressure levels (\url{https://cds.climate.copernicus.eu/datasets/reanalysis-era5-pressure-levels?tab=overview}) for the same duration of 1990-2010. Here, six-hourly data (four times daily) are averaged to obtain daily data. We have integrated the moisture data from $1000$ to $250$ hPa to obtain the column moisture.
The net thermal and solar radiation at both the top of the atmosphere and at the surface, surface sensible heat flux data, and precipitation used to support the WTG approximation are openly available in ERA5 hourly data on single levels at \url{https://cds.climate.copernicus.eu/datasets/reanalysis-era5-single-levels?tab=download}, reference number DOI: 10.24381/cds.adbb2d47. Again, we have averaged the six-hourly data to get the daily mean.
The divergence is obtained from the horizontal winds using the \textit{windspharm} package \citep{dawson2016windspharm}, averaged from 1000 hPa to 500 hPa (middle-lower troposphere), it is used for the graphs of divergence versus column water. The filters used to isolate the westward QBWO and seasonal mean signals remain the same as mentioned in \citep{biswas2026impact}, also determining Day 0 of the composite follows the same method applied there.

\section{Vorticity and Moisture Equations for the QBWO}

We formulate equations for the development of the QBWO in terms of vorticity for the mid to lower troposphere (i.e., from about 400 mbar to 850 mbar) where the system has an upright barotropic structure \citep{chen2010characteristics,biswas2026impact}\citep[for a similar structure of equatorial Rossby waves, see,][]{kw1995,nakamura2022convective} and column moisture. With regard to vorticity, it was noted that, during the boreal summer, the budget in all tropical regions considered takes the form \citep{biswas2026impact},
\begin{align}
    \partial_t \zeta' + (\mathbf{u}\!\cdot \!\nabla \zeta)' + f \nabla \!\cdot \!\mathbf{u}' \approx 0,
    \label{vorticity_a}
\end{align}
where $\mathbf{u}$ is the horizontal flow, $\zeta$ is the vertical component of vorticity, $f$ is the Coriolis parameter and primes refer to QBWO anomalies. In essence, the tendency of the QBWO vorticity anomaly was well approximated by horizontal advection and planetary stretching. To a large extent, horizontal advection (which includes the $\beta$-term) accounts for the propagation of the vorticity anomaly which is the dominant signature in its tendency. Further, across the tropics, from drier to very moist regions, the eddy-eddy term was small and so was $u'\partial_x\bar{\zeta}$ \citep{biswas2026impact}, thus Equation \ref{vorticity_a} simplifies to,
\begin{align}
    \partial_t \zeta' + \bar{\mathbf{u}}\!\cdot \!\nabla \zeta' +  {v}'  \partial_y \bar{\zeta} + \beta v' + f \nabla \!\cdot \!\mathbf{u}' = 0,
    \label{vorticity_b}
\end{align}
where $()'$ and $\bar{()}$ represent QBWO anomalies and the background state, respectively. We have separated out the $\beta$-term and in effect we have the meridional eddy advection of background vorticity as well as the mean flow advection of relative vorticity anomaly. In the context of the QBWO and convectively coupled equatorial Rossby waves, this is in accord with previous work that recognized the importance of the mean flow advecting vorticity anomalies in the lower-mid troposphere for propagation \cite{chen2022model, dong2024propagation}, and the stretching term for growth \citep{matthews2025vorticity}. In fact, the balance between planetary stretching, mean flow advection of vorticity anomalies and the $\beta$-term was prominent in the steady state Rossby wave response to equatorial forcing in the presence of a zonal jet \citep{mont}. %The specific components will be explicitly defined later where we deal with each of the regions in detail. %Though, one thing that is noted in all these regions is that $\bar{\zeta}_x$ is small and the $u'\partial_x\bar{\zeta}$ term can be neglected in Equation \ref{vorticity_b}. %At present we continue with the general formalism that includes all of the possible terms as per Equation \ref{vorticity_b}.

For moisture, the basic conservation principle in pressure coordinates reads,
\begin{align}
    \partial_t q + {\bf u}\cdot \nabla q + \omega \partial_pq = S,
\end{align}
where ${\bf u}$ is again the horizontal flow, $q$ is the moisture, $\omega$ is the pressure velocity and $S$ denotes the sources and sinks of moisture. Denoting the integral over the depth of the troposphere by $\langle \rangle$, we have,
\begin{align}
    \partial_t \langle q\rangle + \langle {\bf u}\cdot \nabla q \rangle = \langle S \rangle - \langle \omega \partial_pq \rangle.
    \label{moisture_a}
\end{align}
Integrating the second term on the RHS by parts and using continuity, the above can also be expressed as,
\begin{align}
    \partial_t \langle q\rangle + \langle {\bf u}\cdot \nabla q \rangle + \langle q \nabla \!\cdot\! {\bf u} \rangle = E-P + Bd,
\end{align}
where $Bd$ is the boundary term in the integral, and $E,P$ are evaporation and precipitation in the column. Of course, the second and third terms on the LHS can be combined to yield the divergence of a flux of moisture (i.e., of ${\mathbf u}q$). Thus, the change in moisture can be due to horizontal advection, convergence (or vertical advection) and a combination of evaporation and precipitation. %Usually, the vertical advection is computed as in Equation \ref{moisture_a}, and this along with precipitation and evaporation is together referred to as the column process \citep{yanai}. 

\begin{figure}[ht]
    \centering
    \includegraphics[width=0.75\linewidth]{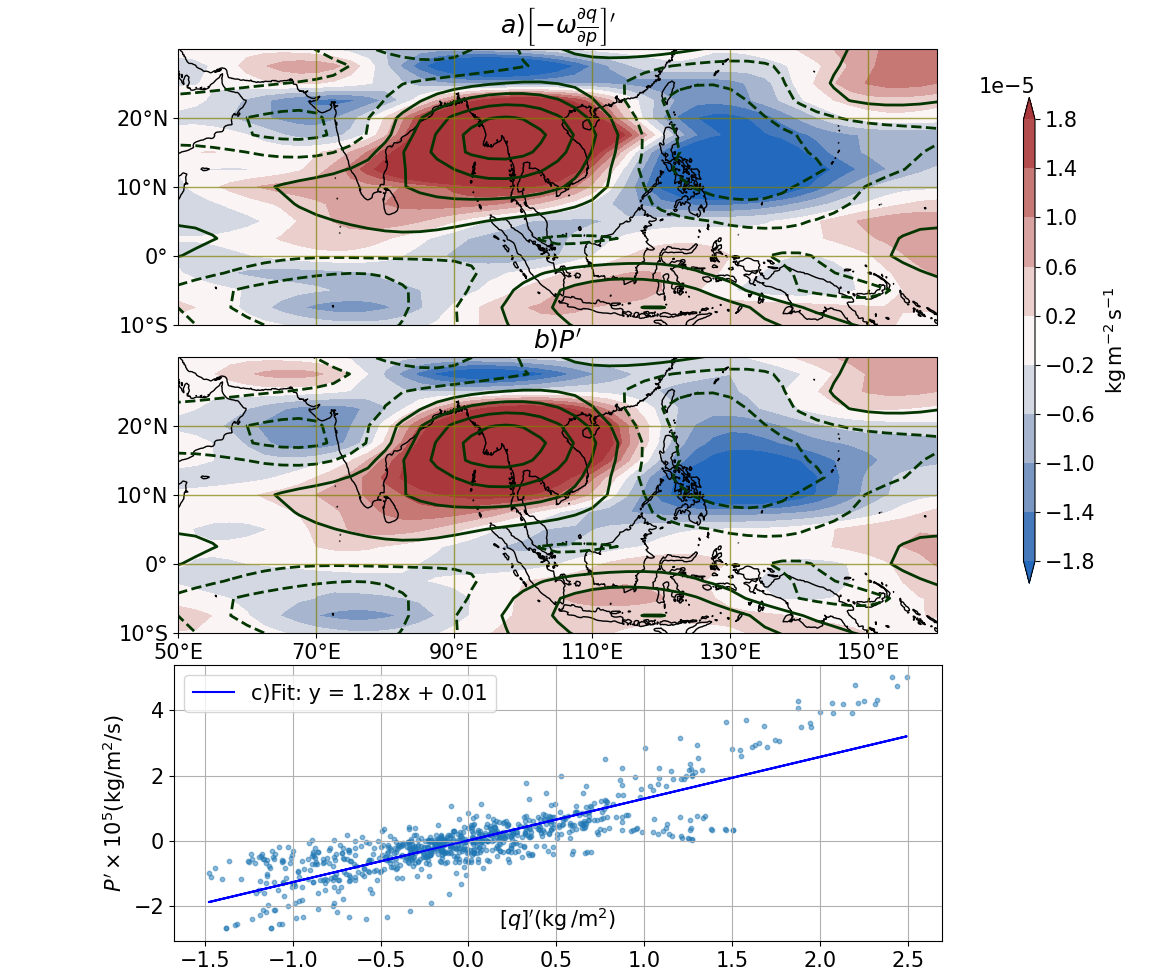}
    \caption{Collocation of column moisture anomaly (contours, intervals of $0.6\;\mathrm{kg/m^2}$ with vertical advection (first row) and precipitation (second row) on Day 0. The third row shows scatterplot and linear fit between precipitation and column moisture anomaly in the map of first and second rows. Here we have used $90^\circ\mathrm{E}-100^\circ\mathrm{E},10^\circ\mathrm{N}-20^\circ\mathrm{N}$ for composite.}
    \label{fig:Figure1}
\end{figure}

In our calculations for the QBWO, in all regions, we have found that the boundary term ($Bd$) is small when integrated from 1000 mbar to about 250 mbar, and there is a very good match between the vertical advection and convergence formalisms (see Figure S1 which shows the match between vertical advection and horizontal convergence in representative dry and moist regions).  Moreover, in terms of QBWO anomalies, evaporation is small and as for most large-scale tropical systems, $P' \propto q'$ \citep{BPB2004,rushley} --- see \cite{nakamura2022bconvective}, for a discussion of this issue in the context of equatorial Rossby waves. Moreover, as seen in Figure \ref{fig:Figure1}, in the very moist regions of the tropics, both $P'$ and $\langle q \nabla \cdot \mathbf{u} \rangle'$ (or $\langle \omega \partial_pq \rangle'$) are aligned with and proportional to $q'$. %\textcolor{blue}{Make a figure showing this for the very moist regions. Merge with the linear relation between P' and q' in these regions. Move this to the main text.} \textcolor{olive}{Figure 1 is inserted.}), %\textcolor{blue}{Make a figure that shows the colocation of these terms in dry and moist regions}) \textcolor{olive}{Supplementary QBWO Model.tex, the maps are attached, for moist regions the relation holds good but for dry region it does not. However, in the dry case, we do not require the $\alpha \langle q\rangle ^\prime$ term.}, %\textcolor{blue}{Show that P'
% and vertical advection or convergence are collocated with q' in a figure. Maybe do a correlation or scatter plot of these quantities.} \textcolor{olive}{Follows for moist region, not valid for dry regions, plots are attached.} 
This leads to a concise form for the column moisture anomaly equation at intraseasonal scales, namely,
\begin{align}
    \partial_t \langle q \rangle' + \langle \mathbf{u}\cdot \!\nabla q \rangle' = -\alpha \langle q \rangle',
    \label{moistue_b}
\end{align}
where, $\alpha$ is a constant that represents the rate at which an existing moisture anomaly is removed from the column in the context of this ``parameterization" --- see, for example, the discussion in \cite{sugiyama-2009,fuchs2019simple,ahmed-neelin}.  Indeed, in terms of moist static energy, the damping of anomalies for intraseasonal Rossby waves by $\langle \omega \partial_pq \rangle'$ has been noted by \cite{mayta2022westward}. In effect, we have a sink of moisture that is proportional to the moisture anomaly and has a timescale of $\alpha^{-1}$ --- much like a Betts-Miller protocol \citep{Betts}, a formulation which has been widely used in shallow water, moist two-layer and idealized general circulation models \citep{sobel2001weak,adames2021,frier}. Note that for relatively dry regions $\alpha \approx 0$ and the balance in the moisture budget is between the tendency and advection \citep{biswas2026impact}. 

Here too, as with the vorticity tendency, the eddy-eddy terms in Equation \ref{moistue_b} are small. In very moist regions with large background moisture gradients, we find the largest contributor to be the eddy advection of background moisture, a feature also observed by \cite{yangSH}. While in other tropical zones, mean flow advection of moisture anomalies dominates the tendency, as has been noted in equatorial Rossby-like moisture modes in the tropical western hemisphere \cite{mayta2022westward} and over the western Pacific Ocean \citep{gonzalez2019distinct}. Thus, across the tropics, a simplified form of the column QBWO moisture anomaly equation reads,
\begin{align}
    \partial_t \langle q \rangle' + \langle \bar{\mathbf{u}}\cdot \nabla q'\rangle + \langle \mathbf{u}'\cdot \nabla \bar q\rangle = -\alpha \langle q \rangle',
    \label{moistue_c}
\end{align}
where $()'$ and $\bar{()}$ represent QBWO anomalies and the background state, respectively.

\begin{figure}[hp]
    \centering
    \includegraphics[width=0.98\linewidth]{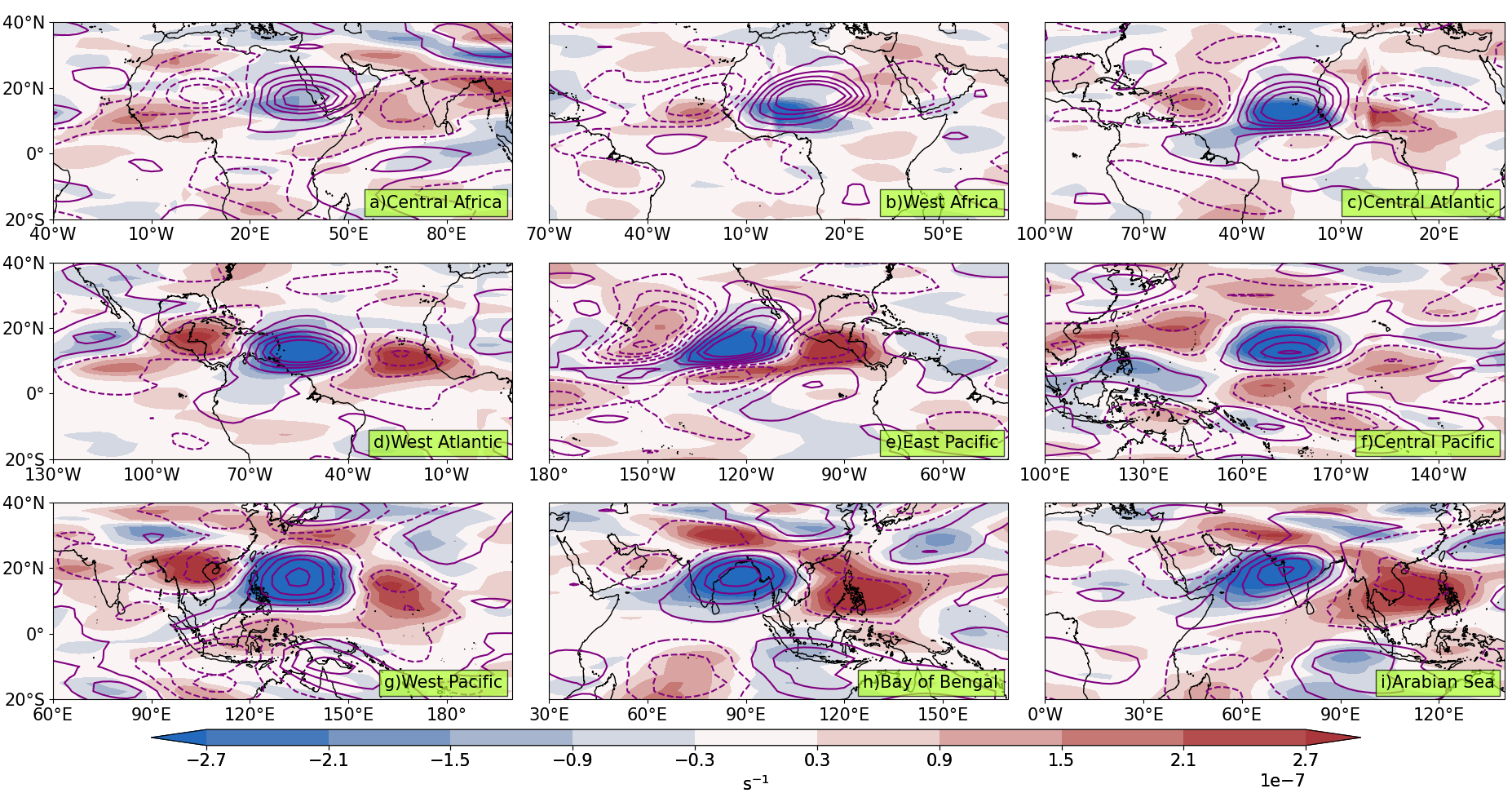}
    \caption{Association of lower tropospheric divergence ($1000\,\mathrm{hPa}$ to $500\,\mathrm{hPa}$, shading) with the column water (contours) on Day 0. The contour intervals are $0.6\,\mathrm{kg/m^2}$.}
    \label{fig:Figure1b}
\end{figure}

\begin{figure}[hp]
    \centering
    \includegraphics[width=0.98\textwidth]{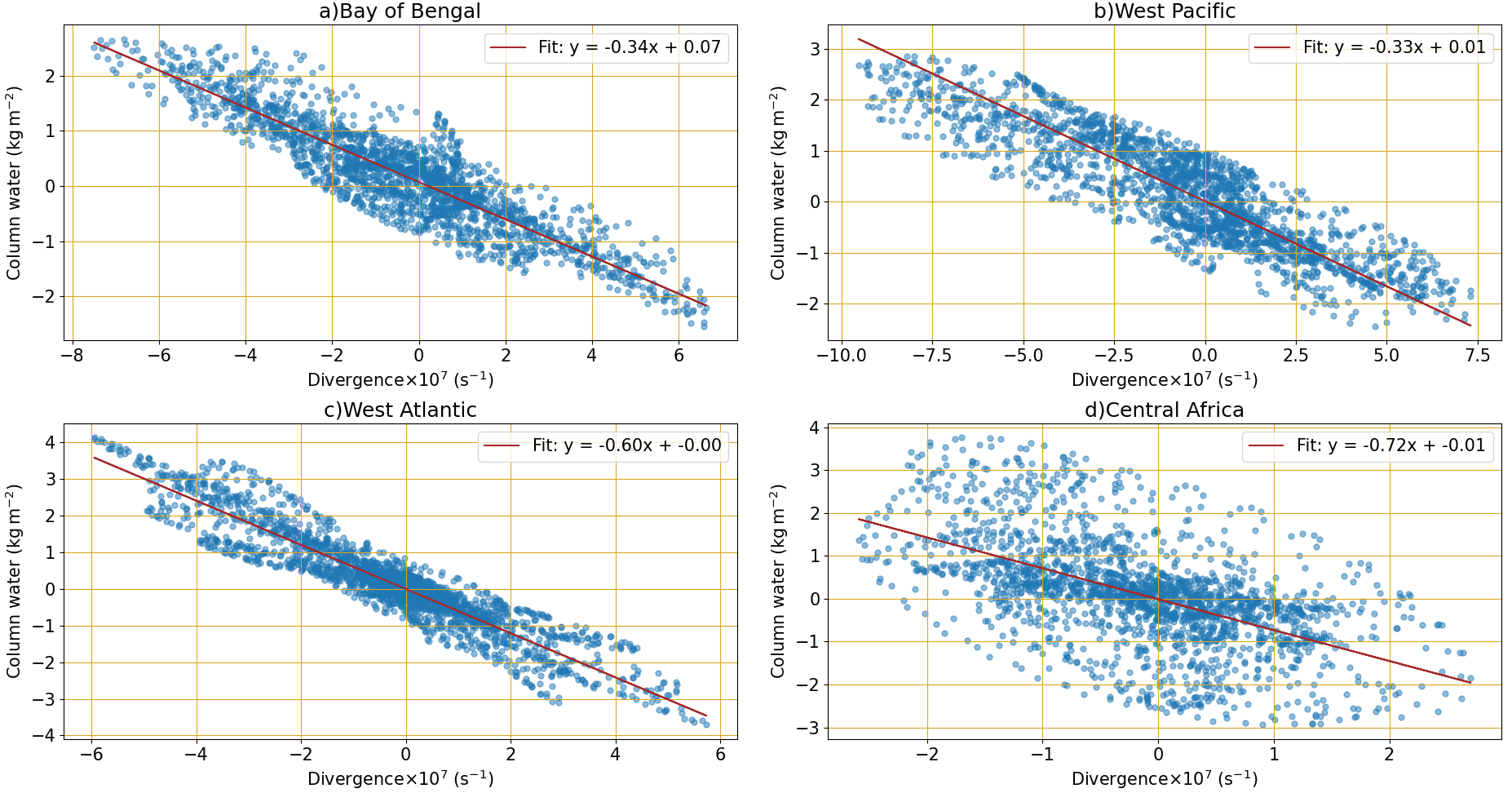}
    \caption{Relation between column moisture and divergence (averaged over $1000-500\;\mathrm{hPa}$) in Central Africa ($5^\circ\mathrm{N-25^\circ\mathrm{N}},0^\circ-50^\circ\mathrm{E}$), Bay of Bengal ($5^\circ\mathrm{N-25^\circ\mathrm{N}},60^\circ-110^\circ\mathrm{E}$), West Pacific ($5^\circ\mathrm{N-25^\circ\mathrm{N}},110^\circ\mathrm{E}-160^\circ\mathrm{E}$) and West Atlantic ($5^\circ\mathrm{N-25^\circ\mathrm{N}},80^\circ\mathrm{W}-30^\circ\mathrm{W}$) regions. The daily variables are chosen on composite Days -5 to +5  in the respective regions.} %This crude relation forms the foundation for closing our system of equations. While the relationship can be assumed to be far complicated, mathematically/computationally we can always close the system  of equations.}
    \label{fig:Figure1a}
\end{figure}

In a general sense, to close this system of equations, we require one equation connecting moisture with dynamics, e.g., the evolution of temperature and divergence. But, as with many large scale, slow tropical modes \citep{adames2022}, the QBWO is in weak temperature gradient (WTG) balance \citep[Figure S2;][]{sobel2001weak,majda-k}. %\textcolor{blue}{Merge current figures 2,3 in supplementary to make one figure.} \textcolor{olive}{Figure S2 is reconstructed.} %\textcolor{olive}{Figures supporting the WTG balance are added in supplementary document.} 
Thus, vertical motion balances diabatic heating, and this column heating is proportional to the moisture anomaly \citep{sobel2001weak,adames2021,ahmed-neelin}. Moreover, given the barotropic nature of the system in the layer under consideration, using continuity, this yields, $\nabla^2\chi' \propto q'$, where $\chi'$ is the velocity potential associated with the QBWO. In fact, empirically, as observed in \cite{biswas2026impact}, the outgoing longwave radiation, moisture, pressure velocity and convergence anomalies are co-located in the QBWO \citep[similar to the convectively active phase of equatorial Rossby waves,][]{nakamura2022convective}. This leads to the closure, 
\begin{align}
    \nabla^2 \chi' = -\lambda q',
    \label{closure}
\end{align}
where $\lambda >0$, which implies enhanced (suppressed) moisture coexists with convergence (divergence), this is evident from the maps in Figure \ref{fig:Figure1b}. The fact that a simple linear relation, with moderate regional variations in the slope, between convergence and the column integrated moisture anomalies holds approximately for the QBWO in different tropical regions can be seen in Figure \ref{fig:Figure1a}. %The column-integrated water-vapour anomalies and mid to lower tropospheric horizontal divergence exhibit an almost linear spatial association across most of the tropical regions examined, with only moderate regional variations in the slope.
Of course, one could fit a more complicated relationship to this data which would still enable a link between these two variables. Due to the WTG balance, the QBWO lives on the slow manifold where gravity waves have been filtered out, and the closure above reduces the three-variable $(\psi',\chi',q')$ system to a slow two-variable problem as $\chi'$ is slaved to $q'$. %In essence, we have a closed system comprising of Equations \ref{vorticity_b}, \ref{moistue_b} and \ref{closure}. %Indeed, this is the form we solve in the following sections for moist and relatively dry tropical regions where different terms dominate the horizontal advection of moisture and vorticity.

\subsection{Non-dimensionalization}

To start our analysis, it is useful to non-dimensionalize the vorticity and moisture anomaly equations. Specifically, we non-dimensionalize Equations \ref{vorticity_b} and \ref{moistue_c} using the closure in Equation \ref{closure}. The scale parameters for non-dimensionalizing the variables are based on the observation that the frequency of the QBWO is comparable to the Rossby frequency ($\omega_R=k\beta_{eff}/K^2$, where $K^2=k^2+l^2$), hence $T=1/\omega_R$. Here $\beta_{eff}=\beta + \partial_y \bar{\zeta}$. The zonal and meridional scales of the eddy are comparable ($l \sim k$), and we set $L\sim 1/k$ and $U \sim L/T \sim \omega_R/k$. The scale parameter ($\mathcal{Q}$) for the moisture anomaly comes from closure, i.e., $\mathcal{Q} \sim \omega_\chi/\lambda$, where $1/\omega_\chi$ is the timescale associated with convergence or stretching. The mean zonal and meridional flows are characterized by a scale $U_0$ and $V_0$, respectively. Finally, as the vorticity anomaly also shares the Rossby frequency as the scale parameter, we obtain,
\begin{align}
&    \frac{\partial \zeta^*}{\partial t} + \left ( \frac{U_0k}{\omega_R} \right) \bar u^* \frac{\partial \zeta^*}{\partial x} + \left ( \frac{V_0l}{\omega_R} \right) \bar v^* \frac{\partial \zeta^*}{\partial y} + \mathcal{B} v^* = \frac{f}{\omega_R}\frac{\omega_\chi}{\omega_R} q^*,  \nonumber \\
&    \frac{\partial q^*}{\partial t} + \left ( \frac{U_0k}{\omega_R } \right) \bar u^* \frac{\partial q^*}{\partial x} + \left ( \frac{V_0l}{\omega_R} \right) \bar v^* \frac{\partial q^*}{\partial y} + \mathcal{M} [\mathbf{u}^* \!\cdot \!(\nabla \bar q_M)^*] = -\frac{1}{\mathcal{N}_\alpha}q^*.
\label{nondim_a}
\end{align}
Here, $()^*$ is used to denote non-dimensional variables and $\mathcal{B}=\beta_{eff}/(k\omega_R) = K^2/k^2 \sim 1$ is a geometric anisotropy factor. Note that the eddy velocity has rotational and divergent components, so here $\mathbf{u}^* = \mathbf{u}^*_\psi + (\omega_\chi/\omega_R) \mathbf{u}^*_\chi$. There are a variety of non-dimensional numbers in Equations \ref{nondim_a}: specifically, $F=f/\omega_R, r_\chi=\omega_\chi/\omega_R, \mathcal{U}=U_0k/\omega_R, \mathcal{V}=V_0l/\omega_R, \mathcal{N}_\alpha = \omega_R/\alpha$ which measure inertial, stretching, mean flow advection, and damping timescales to the Rossby timescale. Finally, we have $\mathcal{M}= \lambda |\nabla \bar q_M|/(\omega_\chi k)$, where $|\nabla \bar q_M|$ is the scale parameter for the gradient of the background moisture field, i.e., we do not take this to be the same as the moisture anomaly. This measures the eddy advection of background moisture relative to the convergence anomaly. An appropriate parameter to measure moisture coupling to Rossby propagation, or moist to dry dynamics, is $\mu=\lambda |\nabla \bar q_M|/\beta_{eff}$, which can be linked to $\mathcal{M}$ via $\mu=r_\chi \mathcal{M}/\mathcal{B}$. We naturally have weak and strong moisture coupling limits comparing the background moisture gradient to $\beta_{eff}$, the effective planetary vorticity gradient, where $\mu \ll 1$ and $\mu \gg 1$, respectively.

\section{Weak and Strong Coupling Regimes}

We study solutions to Equations \ref{nondim_a} in the weak and strong moisture coupling regimes. Specifically, we assume plane-wave solution of the form $A_D'
=
\tilde{A}_D
\exp\!\left[i(\sigma t - kx - ly)\right], A_D \in \{\psi^*, q^*\}$, and examine the change in the nature of the dispersion relations supported in these two conditions.

\subsection{Special Cases}

Before examining the general situation in Equations \ref{nondim_a}, we look at two special cases that guide our intuition, specifically, (i) $\lambda=0$ and (ii) $\nabla \bar q_M=0$, the first has no moist coupling while the second admits no background moisture gradients. For $\lambda=0 \Rightarrow \chi'=0$, and we have a purely rotational flow. %Assuming plane-wave structures $A_D'
%=
%\tilde{A}_D
%\exp\!\left[i(\sigma t - kx - ly)\right], A_D \in \{\psi^*, q^*\}$, 
This yields the dispersion relation,
\begin{align}
    \sigma = \bar u k + \bar v l - \frac{\beta k}{K^2},
\end{align}
which shows that in this case we obtain the usual Doppler shifted Rossby waves. Moreover, moisture anomalies evolve but are merely advected by this flow. On the other hand, for $\lambda \neq0, \nabla \bar q_M=0$, in addition to the Rossby mode we obtain,
\begin{align}
    \sigma= \bar u k + \bar v l+ i \alpha,
\end{align}
i.e., a decaying moisture mode advected by the mean flow. As the moisture gradients are absent, these two solutions do not couple together. This immediately raises the possibility that in the general case, when $\lambda \neq 0, \nabla \bar q_M \neq 0$, these two solutions can couple to yield a moist Rossby-like solution.

\subsection{The General Case}

Proceeding with the general case, we set $\hat\Omega = \hat \sigma - \mathcal{U} u^* - \mathcal{V}v^*$, %\textcolor{olive}{$\hat\Omega = \hat \sigma - \mathcal{U} \bar{u}^* - \mathcal{V}\bar{v}^*$} 
where $\hat \sigma = \sigma/\omega_R$. Substituting a plane wave structure in Equations \ref{nondim_a} yields,
\begin{align}
    (\hat \Omega + 1)(\hat \Omega - \mu A_q - i \delta) = \mu B_q (-m + iF),
    \label{nondim_disp}
\end{align}
here, $A_q$ and $B_q$ are geometrical factors that determine the alignment of the shape-vector $\mathbf{k}=(k,l)$ with background moisture gradient $\nabla \bar q$. Specifically, $A_q=\mathbf{k}^*\!\cdot \!(\nabla \bar q)^*$ and $B_q=(\mathbf{k}^*\!\times \!(\nabla \bar q)^*)_z$, where $k^*=(1,m)$ with $m=l/k$. As before, $(\nabla \bar q)^*$ is the non-dimensional background moisture gradient, $\mu$ is the moisture coupling relative to $\beta_{eff}$ and $F=f/\omega_R$, $\delta=\alpha/\omega_R=1/N_\alpha$. While the non-dimensional form in Equation \ref{nondim_disp} is useful for analysing various limits, the dimensional form is more physically insightful. This reads,
\begin{align}
    (\Omega K^2 + \beta_{eff} k)[ \Omega K^2 - \lambda (\mathbf{k}\!\cdot \nabla \bar q) - i \alpha K^2] = \lambda (\mathbf{k}\!\times \!\nabla \bar q)_z (-\beta_{eff} l + i f K^2),\label{dim_disp}
\end{align}
where $\Omega = \sigma - k \bar u - l \bar v$. We clearly see the roles of the dot and cross products between $\mathbf{k}$ and $\nabla \bar q$ in that the former arises via advection and plays a role in propagation, while the latter arises from the stretching term and likely controls the growth of the solution.

{\it Regime I:} We first examine the weak coupling limit $\mu\ll 1$ of Equation \ref{nondim_disp}. Noting that $F\gg 1$, we have,
\begin{align}
    (\hat \Omega +1)(\hat \Omega - i \delta) = i \mu B_q F. \label{eq14}
\end{align}
Now, taking $\mu F < 1$, expanding about the Rossby branch, i.e., $\hat \Omega = -1 + \epsilon $ with $\epsilon^2 \ll 1$, we obtain,
\begin{align}
    \hat \Omega = -1 - \frac{\mu F B_q \delta}{1+\delta^2} - i \frac{\mu F B_q}{1+\delta^2}.
\end{align}
Or, in dimensional form, the imaginary and real parts of $\sigma$ read,
\begin{align}
    \sigma_i = -\frac{f\lambda (\mathbf{k}\!\times \!\nabla \bar q)_z}{\beta_{eff}k[1+(\alpha/\omega_R)^2]}, \sigma_r = k \bar u+ l \bar v - \omega_R - |\sigma_i| \frac{\alpha}{\omega_R}.\label{eq16}
\end{align}
Given the convention we have followed for plane wave, $\sigma_i<0$ implies growth, and thus we require $\mathbf{k} \times \nabla \bar q >0$ for instability. Such instabilities due to the relative orientation of the wave and the background moisture have been noted in idealized moist models \citep{sukhatme2014low,joy,ahmed2021} and have been formalized in the framework of the moisture-vortex instability \citep[MVI,][]{adames2018,adames2021}. Moreover, $\alpha$ suppresses the growth of the instability. %Note that in the weak coupling limit, $\sigma_i$ in the moist case reduces to that obtained in the dry regions if $\bar q_M=\bar q_M(y), l \approx k$, i.e., the broad mechanism of growth/decay is common to both types of regions and involves the accumulation of moisture via advection, which then feeds into the stretching term to force vorticity growth. 
The propagation of the wave is also altered by the cross-gradient factor, which might be surprising as the projection of $\nabla \bar q$ on the direction of wavevector was seen to arise from advection. In fact, this dot product enters the balance at a higher order, and if we retain $\mu A_q$ in Equation \ref{nondim_disp}, then we obtain an additional modification to $\sigma_r$ of the form $\lambda(\mathbf{k} \cdot \nabla \bar q)/K^2$. 

\begin{figure}[ht]
    \centering
    \includegraphics[width=0.98\linewidth]{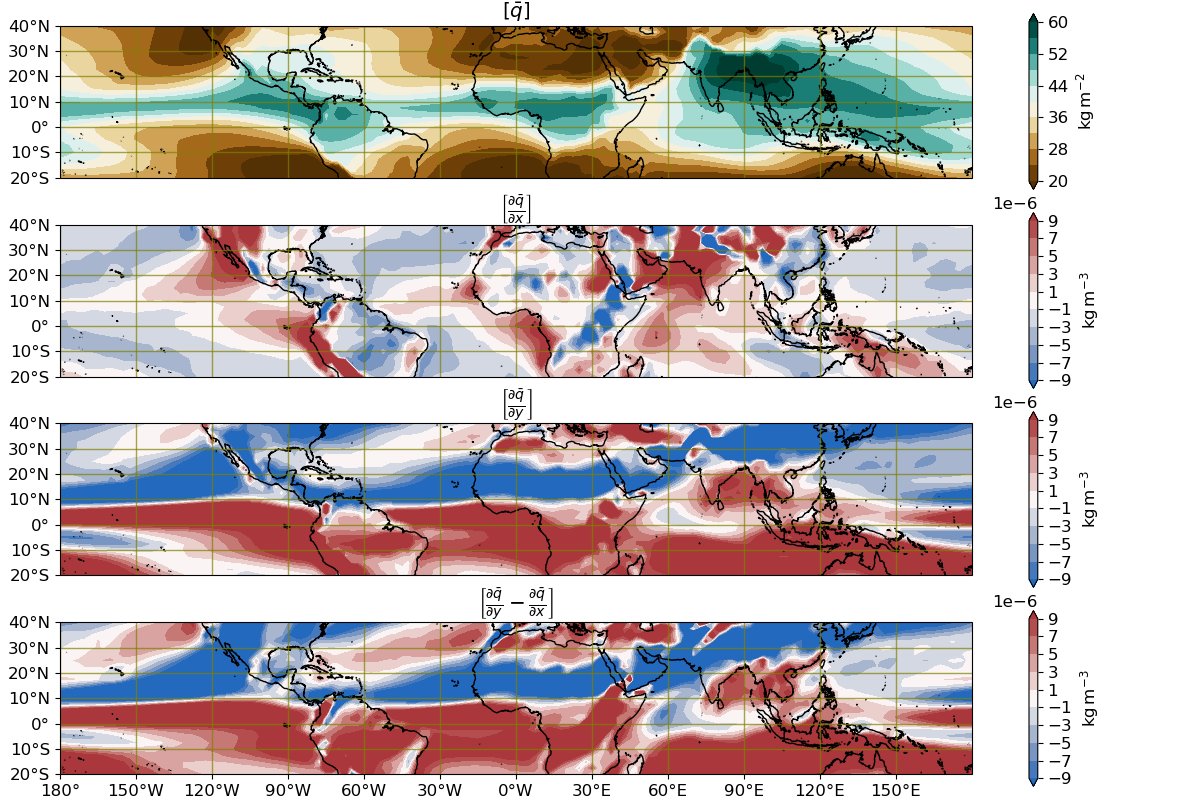}
    \caption{Mean moisture and moisture gradient in JJAS across the tropics. For the growth/decay criteria, $\partial_y\bar{q}-\partial_x\bar{q}$ is also shown, positive (negative) value indicates growth (decay) (for southern hemisphere, the condition is reversed).}
    \label{fig:Figure2}
\end{figure}

Choosing $(k,l)$ to be positive, in the northern hemisphere, growth requires $k \partial_y \bar q - l \partial_x \bar q > 0$, and if $l \sim k$, this implies regions where the background moisture field satisfies $\partial_y \bar q > \partial_x \bar q$ are potential candidates for the growth of this moist Rossby-like mode in the weak coupling regime. In other words, regions flanking maxima of background moisture with $\partial_y \bar q>0$, i.e., a poleward increase in moisture are likely to satisfy this condition. Indeed, a natural candidate, during the boreal summer, is the Bay of Bengal \citep{adames2018}. In addition to this region, a map of $\partial_y \bar q - \partial_x \bar q$ (Figure \ref{fig:Figure2}) for the tropics during the boreal summer shows that the South China Sea is another potential region for the growth of this mode. Moreover, decay is indicated over the subtropical regions of Central and West Africa and many places in the tropical western hemisphere.

\begin{figure}[ht]
    \centering
    \includegraphics[width=0.98\linewidth]{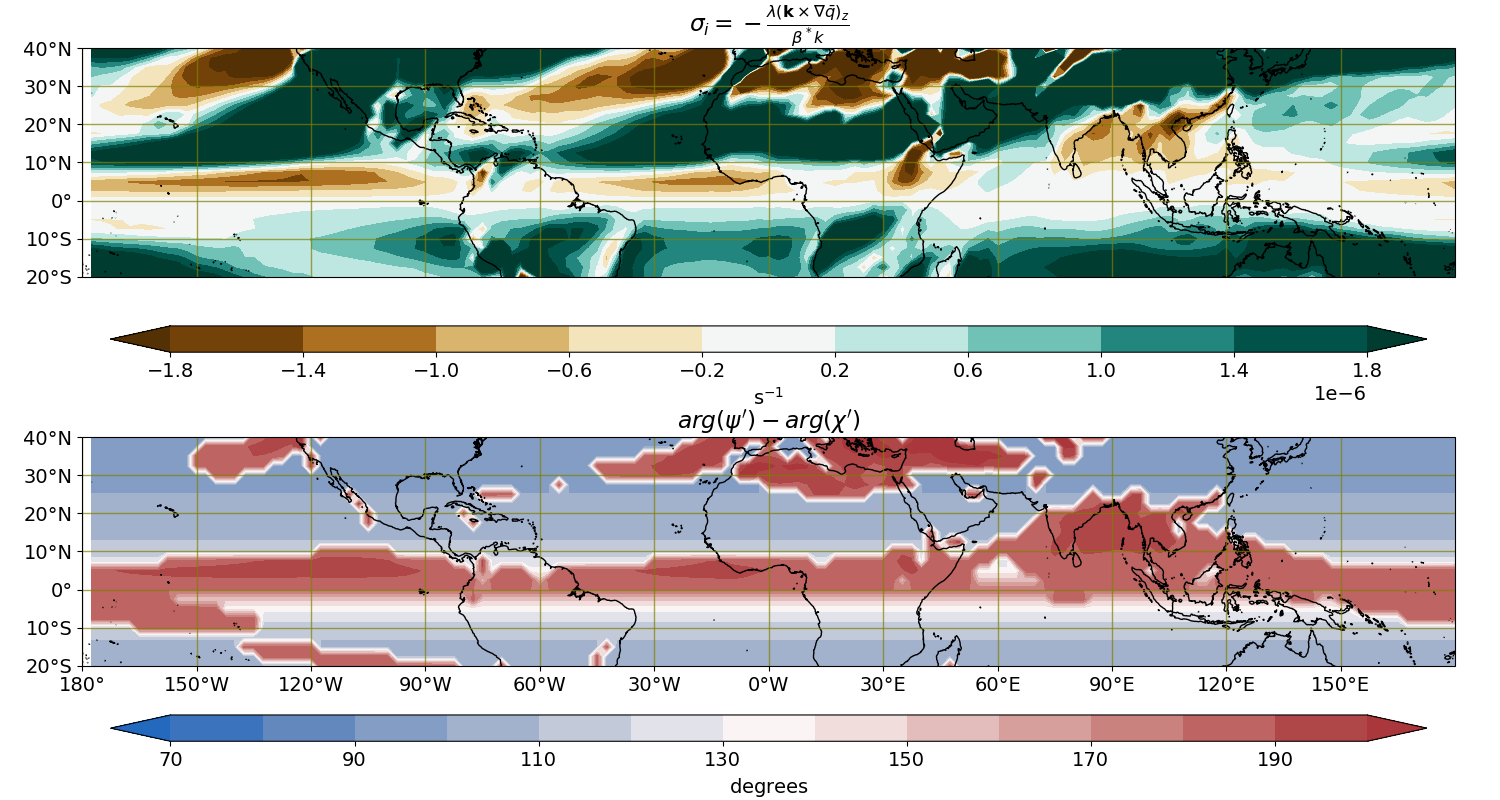}
    \caption{Value of the imaginary part of $\sigma$ and phase difference between $\psi^\prime,\chi^\prime$ according to the applied approximations. For showing the phase difference, $\beta_{eff}=2.21\times 10^{-11}\mathrm{m^{-1}s^{-1}}$ is used throughout and $\alpha=1/10\;\mathrm{day^{-1}}$ is taken for the regions with positive meridional moisture gradient, also we ignored the zonal gradient of moisture in this calculation. Alongside, both numerator and denominator of Equation \ref{eq28} are used inside the range with positive moisture gradient, while only the numerator is used elsewhere to represent the moist and dry regions respectively.}
    \label{fig:Figure3}
\end{figure}
 
{\it Regime II:} The second regime is one of strong moist coupling, i.e., $\mu \gg 1$. Here, Equation \ref{nondim_disp} simplifies to,
\begin{align}
    (\hat \Omega +1)(-A_q) = i F B_q.
\end{align}
therefore,
\begin{align}
    \hat \Omega = -1 - i F \frac{B_q}{A_q}.
\end{align}
Or, in dimensional form, the real and imaginary parts of $\sigma$ read,
\begin{align}
    \sigma_i = -f \frac{(\mathbf{k}\!\times \!\nabla \bar q)_z}{\mathbf{k} \!\cdot \nabla \!\bar q} = -f \tan \theta, \sigma_r = \bar u k+ \bar v l - \omega_R,
\end{align}
where $\theta$ is the angle between $\mathbf{k}$ and $\nabla \bar q$. This indicates rapid growth/decay of the system that dominates over the Rossby wave nature. Only if $\nabla\bar{q}$ is nearly 
parallel to ${\mathbf{k}}$, this effect will be negligible. In effect, for strong coupling, the decay is overwhelmed by stretching and
the system is unstable. As there is no role for $\beta_{eff}$, the system is more like a moist vortex rather than a ``moist wave" as in Regime I. With regard to the phase relation between the vorticity and moisture/convergence anomalies, the moisture equation gives,
\begin{align}
    \frac{\hat \chi}{\hat \psi} = \frac{\lambda B_q k|\nabla \bar{q}|}{\Omega K^2 - \lambda A_q k|\nabla \bar{q}| - i \alpha K^2} \underset{\text{$\mu \gg 1$}}{\approx} -\frac{B_q}{A_q} = -\frac{(\mathbf{k} \!\times \!\nabla \bar q)_z}{\mathbf{k} \!\cdot \!\nabla \bar q}.
\end{align}
Therefore, as $B_q/A_q$ is real, in this regime, the vorticity and convergence/moisture anomalies are in phase with the latter lying inside the QBWO gyre. Given that the growth rate is on the inertial scale, this regime is likely to be too fast to be applicable directly to the QBWO. 

%%%%%%%%%%%%%%%%%%%%%%%%%%%%

%%%%%%%%%%%%%%%%%%%%%%%%%%%%%%%

\subsection{The energy budget}

The energy budget yields a complementary picture that clarifies the role of eddies in extracting energy from the background moisture field. Defining the energy of the eddy field as,
\begin{align}
    \mathcal{V} = \frac{1}{2} (u_\psi'^2 + v_\psi'^2) + \frac{1}{2} \gamma q'^2.
\end{align}
We focus on the rotational portion of the eddy kinetic energy, and $\gamma$ scales $q'^2$ to give a measure of the potential energy. From Equation \ref{moistue_c}, the rate of change of potential energy is (details given in Equations \ref{eqS1} at appendix),
\begin{align}
    \frac{d \langle \mathcal{PE} \rangle}{dt} = - \gamma \langle q'\mathbf{u}'\!\cdot \!\nabla \bar q \rangle - \alpha \gamma\langle q'^2\rangle, 
    \label{m4}
\end{align}
where $\langle \rangle$ denotes the average over the phase of the plane wave solutions. Note that the constant mean flow advection drops out of this average. Here, the extraction of energy from the background moisture field is $G_q = - \gamma \langle q'\mathbf{u}'\!\cdot \!\nabla \bar q \rangle$, while the second term on the RHS leads to a decay of the potential energy. Setting $A'_q=\mathbf{u}'\!\cdot \!\nabla \bar q$, we have, $G_q=-\gamma\langle q' A'_q\rangle = (\gamma/2) \mathbb{R}(\hat{q} \hat{A^*_q})$. Assuming that much of the advection is by the rotational part of the flow, $A_q^\prime \approx \mathbf{u}_\psi'\cdot \nabla \bar q = -i (\mathbf{k} \!\times \!\nabla \bar q)_z \hat \psi$. Therefore,
\begin{align}
    G_q = \frac{\gamma}{2} (\mathbf{k} \times \nabla \bar q)_z |\hat{q}||\hat{\psi}| \sin (\phi_q - \phi_\psi).
\end{align}
We see that the cross product is required for eddy advection of moisture (as was noted in $\sigma_i$ in Equation \ref{eq16}), but this is not sufficient; indeed, the phase of $\psi'$ and $q'$ (equivalently $\chi'$ from closure) need to be correct. In fact, most efficient extraction of background moisture, and build up of $\langle \mathcal{PE} \rangle$ will occur when the two are in quadrature, or, $\phi_{\chi,q}  - \phi_\psi = \pi/2$. 

On the other hand, multiplying Equation \ref{vorticity_b} by $\psi'$ and averaging, after using the closure relation, the growth of rotational eddy kinetic energy is given by (details given in Equations \ref{eqS2} to \ref{eqS7} at appendix),
\begin{align}
%    \frac{d \langle \mathcal{KE} \rangle}{dt} = -f \langle \nabla \psi' \!\cdot \!\nabla \chi' \rangle \propto |\tilde{\chi}_M||\tilde{\psi}_M| \cos (\phi_\chi - \phi_\psi).
\frac{d \langle \mathcal{KE} \rangle}{dt} = -f \langle \nabla \psi' \!\cdot \!\nabla \chi' \rangle + \beta_{eff} \langle \psi' v_\chi'\rangle.\label{m5} %\textcolor{olive}{\frac{d \langle \mathcal{KE} \rangle}{dt} = -f \langle \nabla \psi' \!\cdot \!\nabla \chi' \rangle + \beta_{eff} \langle \psi' v_\chi'\rangle?}
\end{align}
Here, the constant mean flow advection term and the rotational part of the meridional velocity in the $\beta$-term drop out on averaging. The ratio of the $\beta_{eff}$ to stretching contributions scales as $\beta_{eff}l/fK^2 \sim 1/F \ll 1$, i.e., in the QBWO, stretching dominates the rotational eddy kinetic energy budget. Evaluating the stretching term,
\begin{align}
    -f\langle \nabla \psi' \!\cdot \!\nabla \chi'\rangle = -\frac{f \lambda}{2} |\hat \psi| |\hat q| \cos (\phi_q - \phi_\psi)
\end{align}
So, the growth of $\langle \mathcal{KE} \rangle$ via stretching is largest when the $\psi'$ and $\chi'$ (or $q'$) are in phase, or, $\phi_{\chi,q} - \phi_\psi = \pi$ %\textcolor{olive}{Should it be $\pi$ or $180^\circ$ because $\cos\pi=-1$, that will make the stretching term positive. In our moist case, the positive moisture coincides with positive vorticity or negative streamfunction.}. 
So, in terms of the dispersion relations, extraction of energy from the background would be most efficient in Regime I (weak coupling), while the conversion to rotational eddy kinetic energy would be effective in Regime II (strong coupling) when the vorticity and divergence are aligned.

\section{Application to various tropical regions}

Based on the dominant balances in the vorticity and column moisture equations across various tropical regions, the vorticity and moisture dynamics can be broadly categorized into a few distinct classes \citep[specifically, see Tables 1 \& 2 of][]{biswas2026impact}. The first class is characterized by the advection of both perturbed vorticity and moisture by the mean wind, which is representative of the relatively dry regions of the tropics --- specifically, the Central Pacific, East Pacific, West Atlantic, Central Atlantic zones. But the other two dry regions, Central and West Africa, differ in that they also require the advection of the background meridional gradient of moisture by the eddy winds. Thus, within the dry regions, we consider these two categories. On the other hand, the second class features the advection of mean vorticity and moisture primarily driven by the eddy wind, typifying the very moist regions consisting of the Bay of Bengal and the Arabian Sea. Finally, a separate analysis is conducted for the West Pacific, which differs in subtle ways from the drier and very moist regions. While numerous permutations involving different budget terms are possible, this formulation concisely captures the majority of the scenarios identified in our preceding observations.

{\it Central Pacific, East Pacific, West Atlantic, Central Atlantic: } We first consider these dry regions where the background wind $(\bar{u},\bar{v})$ advects the perturbed vorticity or moisture. The eddy wind $(u^\prime,v^\prime)$ does not play a role in advecting the mean moisture, also the decaying effect of column moisture anomaly is negligible and $\alpha \approx 0$. This reduces Equation \ref{dim_disp} to $(\Omega K^2 + \beta_{eff} k)\Omega K^2 = 0$, which produces a typical dry Rossby wave, advected by the mean flow. No growth or decay is detected in this case, mainly due to not considering the eddy wind advection on the mean moisture.

{\it Central and West Africa: } As mentioned, the other dry regions, namely, Central Africa and West Africa, also include the advection of mean meridional moisture by eddy wind, in addition to the considered terms in the previous case. Now, Equation \ref{dim_disp} simplifies to,
\begin{align}
    (\Omega K^2 + \beta_{eff} k)[ \Omega K^2 - \lambda l\frac{\partial \bar{q}}{\partial y}] = \lambda k\frac{\partial \bar{q}}{\partial y} (-\beta_{eff} l + i f K^2)
\end{align}
which, for weak coupling, results in,
\begin{align}
\sigma_r=k\bar{u}+l\bar{v}-\omega_R;\sigma_i=-\frac{f\lambda k \partial_y \bar{q}}{\beta_{eff}k}.\label{dry_speed} 
\end{align}
The real part $(\sigma_r)$ or the propagation mechanism is the same as in the other dry regions. From observations, in these regions, the QBWO gyre is centered around latitude $15^\circ \mathrm{N}$, giving $f=3.77\times 10^{-5}\mathrm{s^{-1}},\beta=2.21\times 10^{-11}\mathrm{m^{-1}s^{-1}}$ ($\beta=\beta_{eff}$ here as we neglect the advection of mean vorticity by eddy winds). The background moisture gradient takes an approximate value of $-10^{-5}\mathrm{kg/m^2/m}$. The planetary zonal wavenumber of the system is around 6, while the meridional wavenumber is around 12. Hence we take $k=6/R_E,l=12/R_E$, where $R_E=6.371\times 10^6\mathrm{m}$. $\lambda$ is obtained from the slope of moisture ($q^\prime$) vs divergence anomaly ($\nabla^2\chi^\prime$) scatterplot in Figure \ref{fig:Figure1a}, giving a value of around $10^{-7}\mathrm{m^2/kg/s}$. As $\partial_y \bar q$ is negative in these regions (i.e., a poleward decay of the background moisture, see Figure \ref{fig:Figure2}), we have a decaying moist Rossby-like wave. The decay time scale ($1/\sigma_i$) gives an approximate value of 7 days. We can also estimate the phase difference between the variables $\psi^\prime,\chi^\prime$ from the vorticity equation. This gives,
\begin{align}
\frac{\hat{\psi}}{\hat \chi}=\frac{-l\beta_{eff}+ifK^2}{\sigma K^2+\beta_{eff}k} = \frac{-l/k + if/\omega_R}{\sigma/\omega_R + 1}=\frac{-m+iF}{\sigma/\omega_R+1}.
\label{eq28}
\end{align}
Clearly, when $F \gg 1$, as $m$ is order unity and the denominator is real, the two fields will show a $\pi/2$ phase difference or a quadrature relation. Similarly, when the meridional wavelength is taken large, or $m\ll 1$, the same quadrature relation will emerge for finite values of $F$. Specifically, the vorticity centre is located at $\pi/2$ east (west) of the divergence (convergence) center in this case. In general, $\hat{\psi}$ leads $\hat{\chi}$ by a phase of $\tan^{-1}(\frac{fK^2}{-l\beta_{eff}})$, which works out to be about $0.84\pi/2$ or $76^\circ$, from the values of $f,k,l,\beta_{eff}$ as listed above. %$\frac{fK^2}{-l\beta}=-4\approx \tan (-76^\circ)$ \textcolor{blue}{How have you got this estimate? Have you used the typical values as listed above? If so, mention this here.}. 
Therefore, in the Central and West African regions, the QBWO appears to consist of a slowly decaying moist gyre where vorticity and moisture/convergence anomalies are almost in quadrature. We can also get an idea about the phase speed of the QBWO in these dry equations from the vorticity equation \ref{vorticity_a}. We can write the equation with streamfunction and velocity potential as follows.
\begin{align}
    \left(\sigma-\bar{u}k-\bar{v}l+\frac{\beta k}{K^2}\right)\tilde{\psi}+\left(\frac{\beta l}{K^2}-i f\right)\tilde{\chi}=0\label{vorticity_c}
\end{align}
Assuming a mean easterly wind $\bar{u}=-5\;\mathrm{m/s}$, $\tilde{\chi}\approx -ir_\chi\tilde{\psi}$, from the quadrature phase relation and dominance of vorticity over divergence by a factor of 10, we can obtain a phase speed of $-5.98\;\mathrm{m/s}$ (from the typical values $\beta /K^2=4.98\;\mathrm{m/s},r_\chi f/k=4\;\mathrm{m/s}$), close to the observed value. However, this is only a check to see whether the vorticity equation is consistent when we apply the observed phase relation and magnitudes of streamfunction and velocity potential. From Equation \ref{dry_speed}, the phase speed is obtained as $-9.98\;\mathrm{m/s}$, almost double the observed values.
%\textcolor{blue}{I am not sure which region you are referring to with the vorticity argument, the vorticity and convergence cannot be in phase here as we just showed its in quadrature.} \textcolor{olive}{This is a hypothetical case where equations for the dry regions are valid, but the mean wind is westerly, partly applicable to the West Pacific region. That creates a positive poleward vorticity which increases $\beta_{eff}$ and decreases the phase angle. If not necessary, we can remove it.}

{\it Bay of Bengal and Arabian Sea:} Moving to the very moist regions with large background moisture gradients, the eddy wind plays a dominant role in advecting the mean moisture here. The Equation \ref{dim_disp} retains its form in this case. As the background flow does not play much role in the moisture budget, effectively $\Omega=\sigma$. We take same values as before for $f=3.77\times 10^{-5}\mathrm{s^{-1}},\beta=2.21\times 10^{-11}\mathrm{m^{-1}s^{-1}},k=6/R_E,l=12/R_E$. Due to background westerlies and considering advection of the mean vorticity by the eddy wind, $\frac{\partial \bar{\zeta}}{\partial y}\approx 10^{-11}\mathrm{m^{-1}s^{-1}}$, making $\beta_{eff}=3.21\times 10^{-11}\mathrm{m^{-1}s^{-1}}$ (see Figure S3). Considering $\lambda=10^{-7}\mathrm{m^2kg^{-1}s^{-1}}$ (from the slope of moisture and divergence anomaly scatterplot in Figure \ref{fig:Figure1}) $,\frac{\partial \bar{q}}{\partial y},\frac{\partial \bar{q}}{\partial x}\approx 10^{-5}\mathrm{{kg/m^2}/m}$ in this region --- notably, as seen in Figure \ref{fig:Figure2}, the Bay of Bengal is mainly affected by $\frac{\partial \bar{q}}{\partial y}$, while the Arabian Sea region is influenced by $\frac{\partial \bar{q}}{\partial x}$. This gives $\mu=\frac{\lambda|\nabla\bar{q}|}{\beta_{eff}}\approx 0.031$, suggesting the applicability of the weak coupling regime, i.e., Equation \ref{eq16}. The decay timescale is obtained is $1/\sigma_i\approx -17\;\mathrm{days}$ (note that a negative value indicates growth) for $\alpha^{-1}=0.5\;\mathrm{day}$ \citep{BPB2004}, $-11\;\;\mathrm{days}$ for $\alpha^{-1}=1\;\mathrm{day}$ \citep{mayta2024stirring} and $-10\;\mathrm{days}$ for $\alpha^{-1}=2\;\mathrm{days}$. Further decrease in $\alpha$ keeps the growth timescale around $10\;\mathrm{days}$. %However, the decay timescale approaches 10 days, as we decrease $\alpha$ \textcolor{olive}{Need to consult literature for $\alpha$}. 
%\textcolor{blue}{The value of $\alpha$ as a convective adjustment is usually taken to be 12 hrs (eg. Bretherton et al 2004), but I think we should report the calculations for a range from 12 hrs to 2 days. Adames et al used about 1 day for BoB.}
The phase relation between $\psi^\prime,\chi^\prime$ reads,
\begin{align}
\arg(\tilde{\psi})-\arg(\tilde{\chi})=\tan^{-1}\left(\frac{fK^2}{-\beta_{eff} l}\right)-\tan^{-1}\left(\frac{\sigma_i}{\Re(\omega_R\epsilon)}\right)=\tan^{-1}\left(\frac{fK^2}{-\beta_{eff} l}\right)-\tan^{-1}\left(\frac{\sgn(\sigma_i)\omega_R}{\alpha}\right).\label{eq30}
\end{align}
Here, $\Re(\omega_R\epsilon)=|\sigma_i|\frac{\alpha}{\omega_R}$ from Equation \ref{eq16}.
The first term gives a value of $110^\circ$, the second term has a value of $-50^\circ$ for $\alpha^{-1}=0.5\;\mathrm{day}$, $-67^\circ$ for $\alpha^{-1}=1\;\mathrm{day}$ and $-78^\circ$ for $\alpha^{-1}=2\;\mathrm{days}$. Further decrease of $\alpha$ converges the value to $-90^\circ$. This explains much of the observed togetherness of vorticity-moisture anomalies. Specifically, vorticity and divergence center are separated by $160^\circ,177^\circ,188^\circ$ (close to $180^\circ$  or $\pi$) for $\alpha=2,1,0.5\;\mathrm{day^{-1}}$ respectively, indicating collocation of vorticity and convergence. However, the phase is sensitive to $\alpha$ and taking $\alpha\sim 1.2\;\mathrm{day^{-1}}$ can exactly account for the observed phase relation. Therefore, over the Bay of Bengal and the Arabian Sea regions, the QBWO appears to be a growing intraseasonal mode in the weak coupling regime. Also, the phase speed is similar to the value obtained from the dry theory here $(-\omega_R/k)$ and satisfies the observed value. Despite having strong background westerlies, the unusual moisture gradient and moisture vortex coupling help to maintain the value of the phase speed here.%\textcolor{blue}{You need to explain this in much more detail. Where is $\partial_x \bar q$? How have you gotten $\mu$? $\lambda$? and $\alpha$? Look at the literature, $\alpha$ is usually 12 hrs or so. Why are you getting decay here? Also, please do not write equations in the text, use the align or equation environment.}

{\it West Pacific:} Finally, we examine the West Pacific region, where both vorticity and moisture are advected by the mean winds. In addition, we cannot ignore $\alpha$ as the column process significantly contributes to the moisture budget. Equation \ref{dim_disp} is reduced to, 
\begin{align}
(\Omega K^2 + \beta_{eff} k)[ \Omega K^2 - i \alpha K^2] = 0.
\end{align}
This suggests a typical dry equatorial Rossby wave, modified by the background wind. The alternate mode is decoupled from the Rossby mode and decays over the timescale $1/\alpha$. Thus, this region also behaves similarly to the dry region. The moisture can partially penetrate the gyre due to the background moisture gradient, i.e., the small zonal and meridional background gradient of moisture can partially account for the in-phase relation between vorticity and convergence in this case, as $\alpha\neq 0$ here. We see that $\beta_{eff}\approx\beta$. As $\beta\approx 2.21\times 10^{-11}\mathrm{m^{-1}s^{-1}}$, this gives $\tan^{-1}\left(\frac{fK^2}{-\beta_{eff}l}\right)=104^\circ$. Now, although we have not introduced advection of the mean moisture by the perturbed wind in this region, it is evident from Equation \ref{eq16} and Equation \ref{eq30} that even a small non-zero value of the mean moisture gradient will enable introduction of an additional phase of $\tan^{-1}(\omega_R/\alpha)$, making the total phase difference around $180^\circ$ or $\pi$, similar to the Bay of Bengal or Arabian Sea. %\textcolor{blue}{How does this compare to the observations? What is the lead lag if 110 is the phase difference?}

\begin{figure}[ht]
    \centering
    \includegraphics[width=0.95\linewidth]{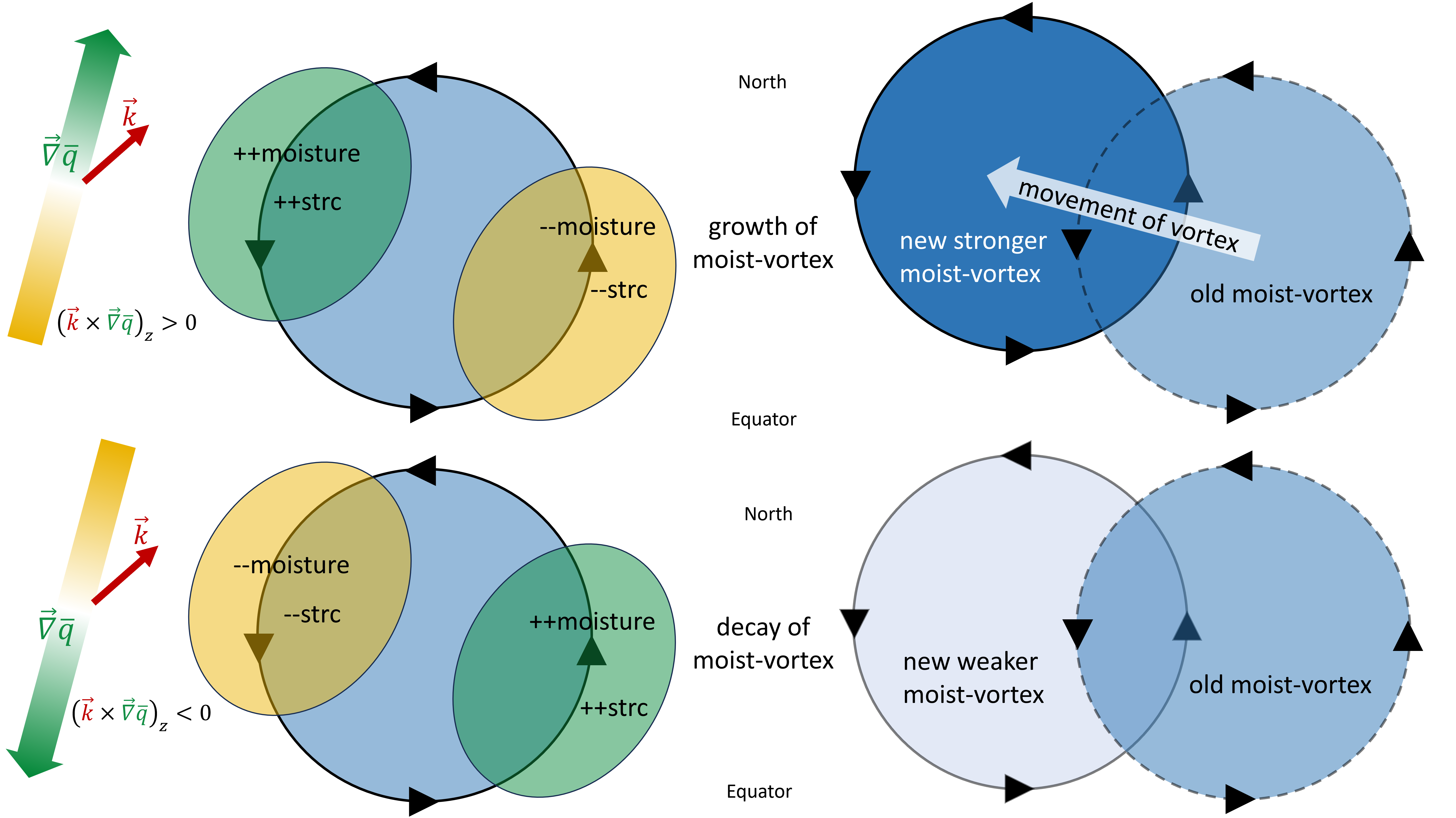}
    \caption{Schematic diagram that explains the structure and propagation of the moisture-vortex in growth and decay configurations for the weak coupling regime (i.e., Equation \ref{eq14}). Growth (decay) occurs if $(\Vec{k}\times\Vec{\nabla}\bar{q})_z$ is positive (negative). The shape-vector $(\mathbf{k}=(k,l))$ is drawn as per the convention used in the manuscript and is directed almost towards north-east. The direction of the moisture gradient, therefore, dictates the instability by inclusion of advected moisture and thereby the stretching term. This immediately illustrates growth (decay) in regions with poleward increase (decrease) in moisture, and is applicable to a large fraction of the tropics.}
    \label{fig:Sch1}
\end{figure}

\section{Discussion and Conclusions}

The Quasi-Biweekly Oscillation (QBWO) is in weak temperature gradient balance (WTG), i.e., its dynamics are constrained to live on the slow manifold defined by this balance, where fast gravity waves that homogenize horizontal temperature fluctuations have been eliminated. Further, as the heating anomalies are largely determined by moisture anomalies, then in essence, this balance reduces the tropical three-variable ($\{\psi',\chi',q'\}$) system to a two-variable system by slaving divergence to moisture anomalies or $\chi' = \mathcal{L}(q')$ --- note that the link between $\chi'$ to $q'$ can be quite general in nature instead of the specific linear closure employed here and in other studies. The physical consistency of the problem, as compared to a prescribed heat source, is that $q'$ evolves (slowly) and thus, so does the divergence and hence the forcing to the vorticity. The coupling comes from the fact that the evolution of $q'$ is tied to the flow, mainly advection by the rotational flow, which is determined from the vorticity equation. In a broader context, as per the classification of \cite{adames2022}, the QBWO is in WTG balance, but its rotational component is not restricted to be in geostrophic, nonlinear or any other balance, i.e., all the dominant terms in the vorticity equation --- depending on the region in question --- are involved in its evolution. %Moreover, following the multiscale perspective of \cite{majda-k}, the QBWO may be viewed as a tropical slow-mode in which convective adjustment diagnoses the divergent circulation, while Rossby-wave dynamics evolve the rotational component. 
Taken together, in line with the multiscale perspective of \cite{majda-k}, the QBWO can be succinctly described as a WTG-constrained moisture–vortex/Rossby mode wherein intraseasonal tropical variability arises from the interaction of convectively adjusted divergent responses with large-scale equatorial Rossby dynamics. %This comprises the dry dynamics of the equatorial Rossby waves \citep{matsuno1966quasi} and provides a venture into the moist regions where the phase difference between moisture and vorticity nears zero and possibility of instability arises.

The instability of the QBWO --- in the weak coupling limit, which seems most physically appropriate to the moist-vortex system given its timescale and the actual value of $\mu=\lambda |\nabla \bar q|/\beta_{eff}$ --- depends on its alignment with the background moisture gradient. A concise illustration of the instability mechanism in Equation \ref{eq14} is presented in Figure \ref{fig:Sch1}, where the moisture-vortex system can grow (decay) on intraseasonal timescales in regions of positive (negative) meridional moisture gradient via involvement of the stretching term. Moreover, as seen in the energy budget analysis, moisture extraction from the background by the QBWO eddies is most efficient when its rotational and divergent components are in quadrature. This results in a build-up of moisture variance. As the moist anomaly grows, the two fields align and vortex stretching results in the growth of the vorticity anomaly and eddy kinetic energy. This is very much in line with previous work that highlighted instabilities due to the nature of the background moisture gradient \citep{sukhatme2014low,joy,suhasIVP,ahmed2021} and the formalized framework of the moisture vortex instability \citep{adames2018,adames2021}. A deficiency of this minimal framework is that the extraction of moisture anomalies in a column (via $\alpha q'$ in Equation \ref{moistue_c}) dampens the growth rate but does not introduce a threshold for instability. This is somewhat counter intuitive in that we anticipate if moisture anomalies are wiped out before they can lead to vortex stretching (via convergence) then the system should be stable --- indeed, this is an issue that bears further investigation. 

The vorticity-moisture framework we have developed works reasonably well across the tropics, from relatively dry to very moist regions. For drier zones, specifically, Central, East Pacific and West \& Central Atlantic regions, the mean winds advect moisture and vorticity anomalies and in some cases the meridional eddy wind advects background moisture (Central and West Africa). In the former, the QBWO is a Doppler shifted Rossby wave, while in the latter, we note a moist Rossby-like solution that decays with a timescale of approximately a week. Moreover, in these regions, as observed, the vorticity and moisture anomalies are nearly in quadrature. On the other hand, for very moist regions (Bay of Bengal and the Arabian Sea), eddy advection of background moisture and vorticity plays a key role, and we find the QBWO to be an unstable moist Rossby-like mode with an intraseasonal growth timescale. Further, in these regions, the moisture and vorticity anomalies are co-located. While encouraging, it should be kept in mind that the timescales noted are dependent on the column moisture decay rate ($\alpha$), which we have chosen to be of the order of one day \citep{BPB2004,mayta2024stirring}. 

The two regimes of weak and strong coupling are strikingly similar to the transition noted in \cite{lap-held}, where moisture first modifies an existing balanced wave instability, but when coupling strengthens, it changes the nature of the dynamics toward vortex amplification. The key difference is the underlying slow manifold. In \cite{lap-held}, the base framework was midlatitude quasi-geostrophic dynamics, and moisture modified baroclinic eddies through latent heating, effective static stability, moist potential vorticity fluxes, and eddy available potential energy-eddy kinetic energy conversion. In this light, the weak and strong coupling limits of our vorticity-moisture coupled system can be viewed as the tropical, WTG-constrained analogue of the moisture-induced wave-to-vortex transition identified in moist baroclinic turbulence \citep{lap-held}.

As a simple model, the presented formalism has a few limitations. First of all, we cannot obtain the height dependence of variables since this is a two-dimensional model. For tilted structures, for example, MJO and Kelvin waves, this model would require modifications. Secondly, the model assumes a beta plane at $15^\circ\mathrm{N}$, which works reasonably well due to the off-equatorial position of the equatorial Rossby waves. Extending this framework to other equatorial waves, such as Kelvin waves and MJO, would require a shift to the equatorial beta plane approximation. %Nevertheless, for a system with significant vorticity and moisture evolution, the scale analysis with the new approximation would produce different solutions in that case. 
Also, we assumed a constant gradient of moisture, but a more realistic structure of the background variables might be required when dealing with smaller scale systems. %However, the methods applied to here can help in improving the existing models and understanding of the dynamics of ER waves from a shallow water perspective. 
With regard to future research, we hope to compare our predictions with outputs from models of intermediate complexity with prescribed ideal conditions. This will help validate the theoretical analysis presented in this paper and further improve our understanding of the main factors influencing the QBWO. Moreover, the observed poleward movement of the QBWO \citep{wang2017quasi, sambrita, yangSH, dong2024propagation} can be explored by further extension of the theory to incorporate zonal gradients of background moisture. %This remains a a possible future direction, especially because the poleward propagation has no direct connection to the westward Rossby wave dynamics.}
 
\clearpage

%\begin{comment}

\begin{comment}

\begin{figure}
    \centering
    \includegraphics[width=0.48\linewidth]{Figures_QBWO_Model/Support_Closure/Moisture_Precipitation_Linear_BoB.png}
    \includegraphics[width=0.48\linewidth]{Figures_QBWO_Model/Support_Closure/Moisture_Vertical_Advection_Linear_BoB.png}
    \caption{Linear relation between moisture anomaly and precipitation (left), vertical advection (right) in the moist region of Indo-China.}
    \label{fig:Figure2}
\end{figure}

\begin{figure}
    \centering
    \includegraphics[width=0.48\linewidth]{Figures_QBWO_Model/Support_Closure/Moisture_Precipitation_Linear_CAf.png}
    \includegraphics[width=0.48\linewidth]{Figures_QBWO_Model/Support_Closure/Moisture_Vertical_Advection_Linear_CAf.png}
    \caption{Weakly linear relation between moisture anomaly and precipitation (left), vertical advection (right) in the dry region of Africa.}
    \label{fig:Figure3}
\end{figure}

\end{comment}

\section*{Acknowledgements}

SB acknowledges support from the Prime Minister’s Research Fellowship (PMRF), the Government of India. JS ac
knowledges support from the Indo-Israel joint collaboration (DST/INT/ISR/P-40/2023). BG acknowledges Anusandhan National Research Foundation (ANRF) research grant SPR/2020/000531. Both SB and JS would like to acknowledge discussions with Prof. Nili Harnik (Tel Aviv University). SB also acknowledges discussions with Prof. Peter Haynes
at the University of Cambridge during his stay as a David Crighton Fellow in the Department of Applied Mathematics
and Theoretical Physics. %The authors also acknowledge the use of ChatGPT5 for pertinent suggestions on the theoretical framework and improving the language in
%the manuscript.
\section*{Conflict of interest}
The authors declare no conflict of interest.

\section*{Data availability statement}
To support and compare with the theoretical framework of this paper, we have inserted values from the following data sources. 
 The outgoing longwave radiation data that support the findings of this study are openly available in NOAA Interpolated Outgoing Longwave Radiation (OLR) at \url{https://downloads.psl.noaa.gov/Datasets/interp_OLR/}, reference Liebmann and Smith, Bulletin of the American Meteorological Society, 77, 1275-1277, June 1996.
The wind, moisture, geopotential, temperature data that support the findings of this study are openly available in ERA5 hourly data on pressure levels from 1940 to present at \url{https://cds.climate.copernicus.eu/datasets/reanalysis-era5-pressure-levels?tab=download}, reference number DOI: 10.24381/cds.bd0915c6 .
The net thermal and solar radiation at both top of the atmosphere and at surface, surface sensible heat flux data, precipitation that support the findings of this study are openly available in ERA5 hourly data on single levels from 1940 to present at \url{https://cds.climate.copernicus.eu/datasets/reanalysis-era5-single-levels?tab=download}, reference number DOI: 10.24381/cds.adbb2d47 .

\bibliography{sample_qbwo_model}

\section{Appendix}
\subsection {Dispersion and phase relations}
The general forms of dispersion relation and phase relation for our two variable system are derived below. Substituting a plane wave solution for streamfunction and velocity potential, we obtain the general form
\begin{align}
    \begin{pmatrix}
        a_1& a_2\\
        b_1& b_2
    \end{pmatrix}
    \begin{pmatrix}
        \tilde{\psi}\\
        \tilde{\chi}
    \end{pmatrix}=\sigma\begin{pmatrix}
        \tilde{\psi}\\
        \tilde{\chi}
    \end{pmatrix}
    \Rightarrow (a_1-\sigma)(b_2-\sigma)-a_2b_1=0\text{ (for non-trivial cases)}.\label{gen_disp1}
\end{align}
This gives us a quadratic equation in $\sigma$. $a_1,a_2,b_1,b_2$ are functions of $(k,l)$ in general, thereby producing the dispersion relation. Also, we can use either the first or second row of \ref{gen_disp1} to derive the phase relation between $\tilde{\psi},\tilde{\chi}$; considering both rows, respectively, gives the following.
\begin{align}
    \frac{\tilde{\psi}}{\tilde{\chi}}=\frac{a_2}{\sigma-a_1}\quad;\quad \frac{\tilde{\psi}}{\tilde{\chi}}=\frac{\sigma-b_2}{b_1}\label{gen_disp2}
\end{align}
The relations are equivalent through the dispersion relation. Therefore, for a particular $\sigma$, we can use any of the stream-function equation or velocity potential equation  to find the phase difference.

\subsection{Energy equations for eddies}
The derivations for the time derivatives of eddy potential energy (\ref{m4}) and eddy kinetic energy (\ref{m5}) are shown here. The potential energy of the eddy field is given by $\mathcal{PE}=\frac{1}{2}\gamma {q^\prime}_M^2$. The rate of change of potential energy is given by
\begin{align}
   \frac{d(\mathcal{PE})}{dt}=\gamma q_M^\prime\frac{dq_M^\prime}{dt}=\gamma q^\prime_M\left(\frac{\partial q_M^\prime}{\partial t}+\bar{\mathbf{u}}\cdot\nabla q_M^\prime+\mathbf{u}^\prime\cdot\nabla\bar{q}_M-\mathbf{u}^\prime\cdot\nabla\bar{q}_M\right)\nonumber\\
   \Rightarrow \frac{d\langle\mathcal{PE}\rangle}{dt}=\gamma \langle q^\prime_M\left(-\alpha q^\prime_M-\mathbf{u}^\prime\cdot\nabla \bar{q}_M\right)\rangle=-\gamma\alpha \langle {q^\prime_M}^2\rangle-\gamma\langle q_M^\prime \mathbf{u}^\prime\cdot\nabla \bar{q}_M\rangle\label{eqS1}
\end{align}
Here, we have used $ \partial_t \langle q \rangle' + \langle \bar{\mathbf{u}}\cdot \nabla q'\rangle + \langle \mathbf{u}'\cdot \nabla \bar q\rangle = -\alpha \langle q \rangle'$. Please note that the potential energy is column integrated because this equation works only for column integrated moisture, assuming the upper and lower boundary of the column have negligible contributions.

The kinetic energy of the eddy field is given by $\mathcal{KE}=\frac{1}{2}\left({u^\prime}_\psi^2+{v^\prime}_\psi^2\right)=\frac{1}{2}|\nabla \psi^\prime|^2$. To calculate the rate of change of $\mathcal{KE}$, we return to the vorticity equation and multiply each term by $\psi^\prime$.
\begin{align}
    \psi^\prime\partial_t \nabla^2\psi' + \psi^\prime(\bar{\mathbf{u}}\!\cdot \!\nabla) \nabla^2\psi'+ \psi^\prime\beta_{eff} v' + \psi^\prime f \nabla^2 \chi^\prime = 0\label{eqS2}
\end{align}
We will average the equation over a complete phase, or equivalently over a wavelength in zonal and meridional directions. We also assume that at the boundaries (denoted by $Bd$), $\psi^\prime$ decays to zero.
Let us take the first term $\langle \psi^\prime\partial_t \partial^2_x\psi'\rangle=\psi^\prime\partial_t \partial _x\psi'|_{Bd}-\langle\partial_x\psi^\prime\partial_t \partial _x\psi'\rangle=-\frac{1}{2}\partial_t\langle\partial_x\psi^\prime \partial _x\psi'\rangle=-\frac{1}{2}\langle\partial_t{(\partial_x\psi^\prime)}^2\rangle$. Similarly $\langle \psi^\prime\partial_t \partial^2_y\psi'\rangle=-\frac{1}{2}\partial_t\langle{(\partial_y\psi^\prime)}^2\rangle$. Together we get
\begin{align}
\langle\psi^\prime\partial_t \nabla^2\psi'\rangle=-\frac{1}{2}\partial_t\langle{|\nabla\psi^\prime|}^2\rangle=-\partial_t\langle\mathcal{KE}\rangle.\label{eqS3}
\end{align}
Now we perform the average over the second term. For simplicity, we evaluate the term $\psi^\prime\bar{u}\partial_x \nabla^2\psi'$ and notice that the $x$ part of the Laplacian gives, $\langle\psi^\prime\bar{u}\partial_x \partial_x^2\psi'\rangle=\psi^\prime\bar{u}\partial_x \partial_x\psi'|_{Bd}-\langle\partial_x\psi^\prime\bar{u}\partial_x \partial_x\psi'\rangle=-\langle\frac{1}{2}\bar{u}\partial_x(\partial_x\psi^\prime\partial_x\psi^\prime)\rangle$. Similarly, $\langle\psi^\prime\bar{u}\partial_x \partial_y^2\psi'\rangle=\psi^\prime\bar{u}\partial_y \partial_x\psi'|_{Bd}-\langle\partial_y\psi^\prime\bar{u}\partial_x \partial_y\psi'\rangle=-\langle\partial_y\psi^\prime\bar{u}\partial_x \partial_y\psi'\rangle=-\langle\frac{1}{2}\bar{u}\partial_x\partial_y\psi^\prime\partial_y\psi^\prime\rangle$ gets a similar form. Together, we can write and extend for the meridional advection term.
\begin{align}
\langle\psi^\prime\bar{u}\partial_x \nabla^2\psi'\rangle=-\frac{1}{2}\langle\bar{u}\partial_x|\nabla\psi^\prime|^2\rangle\quad ; \quad\langle\psi^\prime\bar{v}\partial_y \nabla^2\psi'\rangle=-\frac{1}{2}\langle\bar{v}\partial_y|\nabla\psi^\prime|^2\rangle\nonumber\\
\Rightarrow \langle\psi^\prime(\bar{\mathbf{u}}\!\cdot \!\nabla) \nabla^2\psi'\rangle=-\frac{1}{2}\langle \bar{\mathbf{u}}\cdot\nabla|\psi^\prime|^2\rangle=-\bar{\mathbf{u}}\!\cdot\!\nabla(\mathcal{KE})\label{eqS4}
\end{align}
The $\beta$-term is quite straightforward, here the irrotational velocity remains as non-zero component.
\begin{align}
\langle\psi^\prime\beta_{eff} v'\rangle=\langle\psi^\prime\beta_{eff} (\partial_x\psi+\partial_y\chi)'\rangle=(1/2)\psi^\prime\beta_{eff}\psi^\prime|_{Bd}+\langle\psi^\prime\beta_{eff}v^\prime_\chi\rangle=\beta_{eff}\langle\psi^\prime v^\prime_\chi\rangle\label{eqS5}
\end{align}
For the last term, we can apprehend that $\langle\psi^\prime\partial_x^2\chi^\prime\rangle=\psi^\prime\partial_x\chi^\prime|_{Bd}-\langle\partial_x\psi^\prime\partial_x\chi^\prime\rangle$. Extending to the meridional part of the Laplacian, we get the form below.
\begin{align}
    \langle\psi^\prime f \nabla^2 \chi^\prime\rangle=-f\langle\nabla\psi^\prime\cdot\nabla\chi^\prime\rangle\label{eqS6}
\end{align}
The averaged form of the Equation \ref{eqS2} therefore behaves as follows, we assume that $\mathbf{u}^\prime\nabla\mathcal{KE}$ is negligible.
\begin{align}
    -\partial_t(\langle\mathcal{KE}\rangle)-\bar{\mathbf{u}}\cdot\nabla(\langle\mathcal{KE}\rangle)+\langle\psi^\prime\beta_{eff}v^\prime_\chi\rangle-f\langle\nabla\psi^\prime\cdot\nabla\chi^\prime\rangle=0\nonumber\\
    \Rightarrow \partial_t(\langle\mathcal{KE}\rangle)+\bar{\mathbf{u}}\cdot\nabla(\langle\mathcal{KE}\rangle)\approx\frac{d\langle\mathcal{KE}\rangle}{dt}=\beta_{eff}\langle\psi^\prime v^\prime_\chi\rangle-f\langle\nabla\psi^\prime\cdot\nabla\chi^\prime\rangle\label{eqS7}
\end{align}
The forms \ref{eqS1}, \ref{eqS7} are used for the rate of change of eddy potential and kinetic energy respectively.
\end{document}